\documentclass{article}
\usepackage{styles/preprint/arxiv}
\usepackage{cite}
\usepackage{amsmath,amssymb,amsfonts}
\usepackage{graphicx}
\usepackage{textcomp}

\usepackage[utf8]{inputenc}
\usepackage[english]{babel}
\usepackage[T1]{fontenc}

\usepackage{nicefrac}
\usepackage{siunitx}
\usepackage{booktabs}
\usepackage{caption}
\usepackage{subcaption}
\usepackage{multirow}

\usepackage[pangram]{blindtext}

\renewcommand\footnotemark{}

\begin{document}

\title{Wearable Sensor-based Multimodal Physiological Responses of Socially Anxious Individuals across Social Contexts}

\author{Emma R. Toner*, Mark Rucker*, \\Zhiyuan Wang, Maria A. Larrazabal, Lihua Cai, Debajyoti Datta, Elizabeth Thompson, \\Haroon Lone, Mehdi Boukhechba, Bethany A. Teachman, and Laura E. Barnes \vspace{-1cm}}

\date{}

\thanks{*Emma R. Toner and Mark Rucker contributed equally to the manuscript and are co-first authors.}
\thanks{Research in this publication was partially supported by a 3Cavaliers Seed Grant and by the National Institute of Mental Health of the National Institutes of Health under award number R01MH132138. The content is solely the responsibility of the authors and does not necessarily represent the official views of the National Institutes of Health.}
\thanks{Emma R. Toner, Maria A. Larrazabal, and Bethany A. Teachman are with the Department of Psychology, University of Virginia, Charlottesville, Virginia.} 
\thanks{Mark Rucker, Zhiyuan Wang, Debajyoti Datta, Elizabeth Thompson, and Laura Barnes are with the Department of Systems and Information Engineering, University of Virginia, Charlottesville, Virginia.}
\thanks{Lihua Cai is with the Aberdeen Institute of Data Science and Artificial Intelligence, South China Normal University, Guangzhou, China.}
\thanks{Haroon Lone is with the Department of Electrical Engineering and Computer Science, Indian Institute of Science Education and Research Bhopal, Bhopal, India.}
\thanks{Mehdi Boukhechba is with Janssen Pharmaceutical Companies of Johnson \& Johnson, Titusville, NJ 08560 USA.}

\maketitle

\begin{abstract}

Correctly identifying an individual's social context from passively worn sensors holds promise for delivering just-in-time adaptive interventions (JITAIs) to treat social anxiety disorder. In this study, we present results using passively collected data from a within-subject experiment that assessed physiological response across different social contexts (i.e, alone vs. with others), social phases (i.e., pre- and post-interaction vs. during an interaction), social interaction sizes (i.e., dyadic vs. group interactions), and levels of social threat (i.e., implicit vs. explicit social evaluation). Participants in the study ($N=46$) reported moderate to severe social anxiety symptoms as assessed by the Social Interaction Anxiety Scale ($\geq$34 out of 80). Univariate paired difference tests, multivariate random forest models, and follow-up cluster analyses were used to explore physiological response patterns across different social and non-social contexts. Our results suggest that social context is more reliably distinguishable than social phase, group size, or level of social threat, but that there is considerable variability in physiological response patterns even among these distinguishable contexts. Implications for real-world context detection and deployment of JITAIs are discussed. 
\end{abstract}

\section{Introduction}

Social anxiety disorder (SAD) is among the most common mental health disorders in the United States: an estimated 13\% of adults will be diagnosed with SAD during their lifetime \cite{kessler_twelve-month_2012}. Characterized by intense anxiety about social situations, fear of negative evaluation, and avoidance of social situations \cite{american_psychiatric_association_diagnostic_2022}, SAD is associated with substantial functional impairment across work and social domains \cite{aderka_functional_2012}. Despite the burden conferred by SAD, most individuals with this diagnosis do not seek or cannot access treatment \cite{grant_epidemiology_2005} and many wait decades before receiving care \cite{wang_failure_2005}. If effective psychological treatments exist for SAD \cite{acarturk_psychological_2009}, why do so many people fail to receive treatment? One particularly compelling explanation is that the dominant mode of psychological treatment delivery (i.e., one-on-one therapy delivered in-person by a trained mental health professional) is difficult to access and cannot meet the overwhelming need \cite{kazdin_addressing_2017}. For socially anxious individuals, there are also frequently logistical, financial, and emotional (e.g., shame; anxiety about talking to a provider) barriers to seeking treatment \cite{goetter_barriers_2020}. 

Digital mental health interventions (DMHIs) offer a potential alternative to the standard, one-on-one treatment delivery model as DMHIs are typically accessible, low-cost, and private. In particular, just-in-time adaptive interventions (JITAIs; \cite{nahum-shani_just--time_2018}), which aim to personalize and optimize DMHIs by delivering context-relevant interventions at times when they are most needed, may hold particular promise for closing the social anxiety treatment gap. To most effectively deploy JITAIs, it is necessary to determine when people are most in need of an intervention and which intervention(s) would be most useful and appropriate given their current context (e.g., their location). One way to gather this information is by asking participants to report on their emotional experiences and situational contexts in daily life (e.g., via ecological momentary assessment), but this may be burdensome \cite{van_genugten_experienced_2020} and requires participants to respond to prompts frequently and accurately. Additionally, it may be challenging to collect self-report data when participants are actively engaged in contexts of interest (e.g., social interaction) given that responding to a prompt would mean interrupting whatever they are doing. A potential alternative to directly asking participants about their experience is to passively gather data (e.g., physiological responding; GPS location) via unobtrusive devices and sensors (e.g., wristbands; cell phones) that can help identify one's current emotional state and/or situational context. 

Identifying social \textit{contexts} can critically inform social anxiety interventions given that socially anxious individuals are more likely to experience anxiety in certain situations compared to others (e.g., social vs. non-social or evaluative vs. non-evaluative situations; \cite{heimberg_cognitive-behavioral_2014}). Additionally, some intervention strategies are more or less feasible to implement depending on the context (e.g., if a person is actively engaged in a social interaction, they will likely be unable to write down and evaluate all of their thoughts in detail). Accordingly, understanding \textit{where} someone is or \textit{what} they are doing can aid in the determination of when to deploy an intervention and the selection of an intervention that is likely to be appropriate and effective. 

Advances in mobile technology and passive sensing have made it possible to detect social anxiety and social context in experimental studies and in daily life from biobehavioral data. For instance, physiological data from wearable devices (e.g., skin temperature; skin conductance) has been used to accurately detect social anxiety severity in the context of a speech task \cite{shaukat-jali_detecting_2021}, and other passively sensed data from mobile phones (e.g., GPS; call/text data) has been leveraged to detect symptoms of social anxiety in daily life \cite{rashid_predicting_2020, boukhechba_predicting_2018}. Additionally, physiological biomarkers such as heart rate, skin conductance, skin temperature, vocal pitch, and hand movements can be leveraged to accurately differentiate between baseline, anticipatory, and concurrent phases of social anxiety \cite{shaukat-jali_detecting_2021, pisanski_multimodal_2018} and between levels of evaluative threat during a speech task \cite{barreda-angeles_users_2020}. In daily life studies, audio data has been used to identify when participants are near human speech (e.g., conversations; \cite{wang_studentlife_2014, ben-zeev_next-generation_2015}) and location data can help detect when people are interacting in a group and differentiate between group sizes \cite{zakaria_stressmon_2019}. 

Taken together, these findings suggest that passively-sensed data is an important tool that can be used to determine when an individual may benefit from a targeted intervention. However, many open questions remain. Thus far, most efforts to detect social anxiety or social contexts have emphasized using many biobehavioral features together, which can maximize prediction but reduce the scalability of these efforts (e.g., if participants are required to wear numerous devices to collect all of the data necessary for accurate prediction), and make it hard to determine which features are most needed (and thus should be prioritized when planning data collection) and which may be redundant. 

\begin{figure*}
    \centering
    \includegraphics[width=1\textwidth]{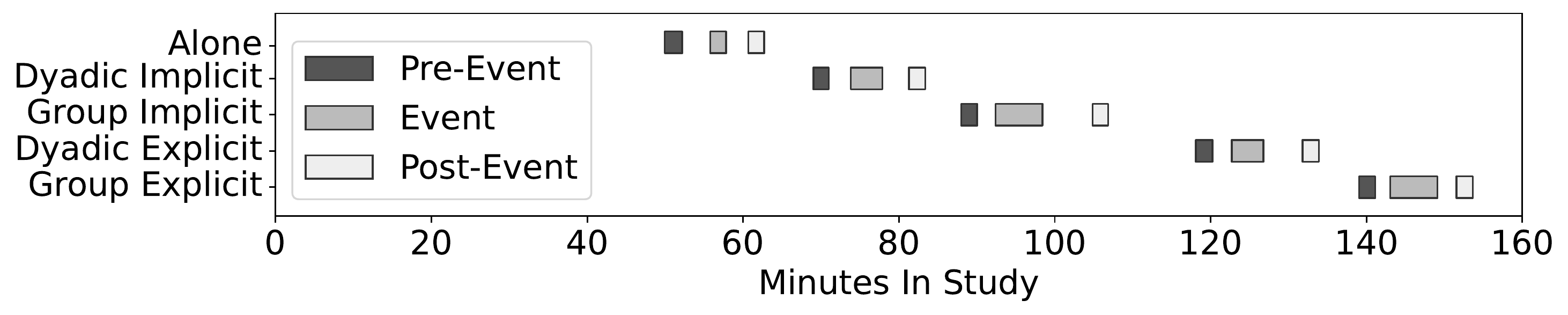}
    \caption{We show a complete experimental session for a participant. The order of events (y-axis) was randomized for each session --- except for Alone, which always came first. The initial 45 minute lag was due to study setup needs.}
    \label{fig:phases}
    \vspace{-.2cm}
\end{figure*}

The present investigation leverages data from an experimental study conducted virtually via Zoom in which individuals high in trait social anxiety completed a non-social task and four social interactions involving different numbers of interaction partners and levels of experimentally-manipulated social-evaluative threat. Physiological indicators of anxiety were passively assessed via an Empatica E4 wristband and we extracted 13 features relevant to anxiety and stress detection based on prior literature (see Table~\ref{tab:feats}). This exploratory study aims to advance our understanding of context-specific physiological responding of socially anxious individuals in three primary ways:
 
\begin{enumerate}
    \item Describe the physiological responses to both social and non-social contexts and phases. 
    \item Determine significantly different physiological responses to both social and non-social contexts and phases.
    \item Explore, via clustering analyses, whether physiological responses can be characterized using a small set of responses or if responses are relatively heterogeneous across social and non-social contexts and phases. 
\end{enumerate}


\section{Method}

\subsection{Participants}
Participants were undergraduate students at a large public university in the Southeastern United States. $N = 55$ participants consented to participate for psychology course credit. All participants completed the Social Interaction Anxiety Scale (SIAS; \cite{mattick_development_1998}); only those scoring 34 or higher, indicating moderate to severe symptoms of social anxiety, were eligible to participate. Nine participants were excluded due to low SIAS scores, yielding a final sample of $N = 46$. The average pre-enrollment SIAS score was 45.57 (\textit{SD} = 8.87), with scores ranging from 34-69 (out of a possible total of 80). Additionally, participants were excluded if they endorsed routine use of certain medications (bendodiazepines, stimulants, antipsychotics, mood stabilizers, beta-blockers, monoamine oxidase inhibitors, medications that cross the blood-brain barrier, or pain medications) or a diagnosis of cardiovascular disease, all of which can influence psychophysiological reactivity. For this reason, participants were also asked to refrain from using the aforementioned medications, along with caffeine, nicotine, and marijuana, and avoid engaging in vigorous exercise for at least two hours prior to the study session, and to avoid alcohol and illegal drugs for at least 24 hours prior to the study session. 

\subsection{Procedure}
All study procedures were approved by a university Institutional Review Board (UVA-SBS IRB \#3004). Participants attended a three-hour study session, which was conducted virtually via Zoom due to the COVID-19 pandemic, in groups of four to six. Sessions were led by trained undergraduate and/or graduate students in psychology. During the study session, participants completed up to four social tasks and one non-social task. The first task completed was always a non-social baseline in which participants watched one of three neutral videos of colorful moving shapes for two minutes by themselves (valence for the the neutral videos was established using a prior mTurk sample; \cite{toner_social_2022}). Subsequently, participants completed two dyadic conversations (lasting four minutes each) and two group conversations (lasting six minutes each) in random order. In each conversation, participants were provided with one of four pre-selected conversation topics (e.g., "If you won a million dollars, what would you do with the money and why?"). For each of the two dyadic and two group conversations, one was designated as being \textit{explicitly} socially evaluative, such that participants were informed prior to the conversation that their conversation partner(s) would rate them on the basis of their likeability and conversation skills after talking. The other two conversations did not involve this instruction and were thus \textit{implicitly} socially evaluative, deemed as such because socially anxious individuals are likely to perceive any unfamiliar social interaction as involving potential social-evaluative threat, even if they are not explicitly made aware of this threat \cite{heimberg_cognitive-behavioral_2014}. All five tasks were also divided into three separate temporal phases: (1) the anticipatory or \textit{pre-event} phase, in which participants were informed about the upcoming task and instructed to vividly imagine doing the task for two minutes (e.g., imagine themselves talking to a stranger who would then be evaluating them); (2) the \textit{concurrent} phase, during which participants completed the task; and (3) the \textit{post-event} phase, in which participants were instructed to spend two minutes vividly reflecting on the task they just completed and how they felt the task went. Participants' physiological responding was continuously monitored throughout the three-hour study session via an Empatica E4 wristband (https://www.empatica.com/research/e4/) worn on their left wrist. See Figure~\ref{fig:phases} for a visual depiction of study procedures. 

\begin{table}
    \centering
    \caption{The four sensors used in this study.}{
    \begin{tabular}{ c c l l c }
        \toprule
        Name & Rate & Description & Location & Resolution \\
        \midrule
        PPG & 64Hz & Photoplethysmography & Outer Wrist  & \SI{0.9}{\nano\watt} \\
        ACC & 32Hz & 3-Axis Accelerometer & Outer Wrist  & \SI{0.016}{g} \\
        EDA & 4Hz & Electrodermal Activity & Inner Wrist & \SI{\sim900}{\pico\siemens}\\
        TMP & 4Hz & Skin Temperature & Outer Wrist       & \SI{0.02}{\degreeCelsius}\\
         \bottomrule
    \end{tabular}}
    \label{tab:sensors}
\end{table}

\subsection{Data}
 For the present study, we focus on data collected from four of the Empatica E4 sensors: photoplethysmography (PPG), accelerometer, electrodermal activity (EDA), and skin temperature. These four sensors are a subset of the entire study dataset, which included data from additional sensors and devices, and were selected due to their alignment with existing works in the literature on passive anxiety detection (\cite{sun2012activity,boukhechba2018demonicsalmon,hernando2018validation,ahn2019novel,barreda2020users,ihmig2020line,petrescu2020integrating,hickey2021smart}). Detailed technical descriptions of the four sensors used in this study can be found in Table~\ref{tab:sensors}.

\subsection{Data Cleaning}
\label{sec:cleaning}

\paragraph{Selection} We began cleaning the data by excluding all sensor readings collected outside of the experimental phases and experiences. This left us with between nine to fifteen (depending on how many tasks a participant completed) distinct intervals of sensor readings for each participant, with a few minutes gaps in-between the intervals (see Figure~\ref{fig:phases}).

\paragraph{PPG} We applied a $3^{rd}$ order Butterworth filter with a low cut of 0.5Hz and a high cut of 8Hz to the raw PPG readings. This removed any artifacts which ocurred at frequencies outside of a normal heart rate. We then used the peak detection algorithm in \cite{elgendi2013systolic} to extract systolic peaks, and calculate the normal-to-normal (NN) intervals. Both the filter and the peak detection algorithm were applied using Neurokit2 \cite{Makowski2021neurokit}. 

\begin{figure*}[t]
\begin{minipage}[t]{.48\linewidth}
\vskip 0pt
{
    \vspace{-.17cm}
    \centering
    \includegraphics[width=.96\textwidth]{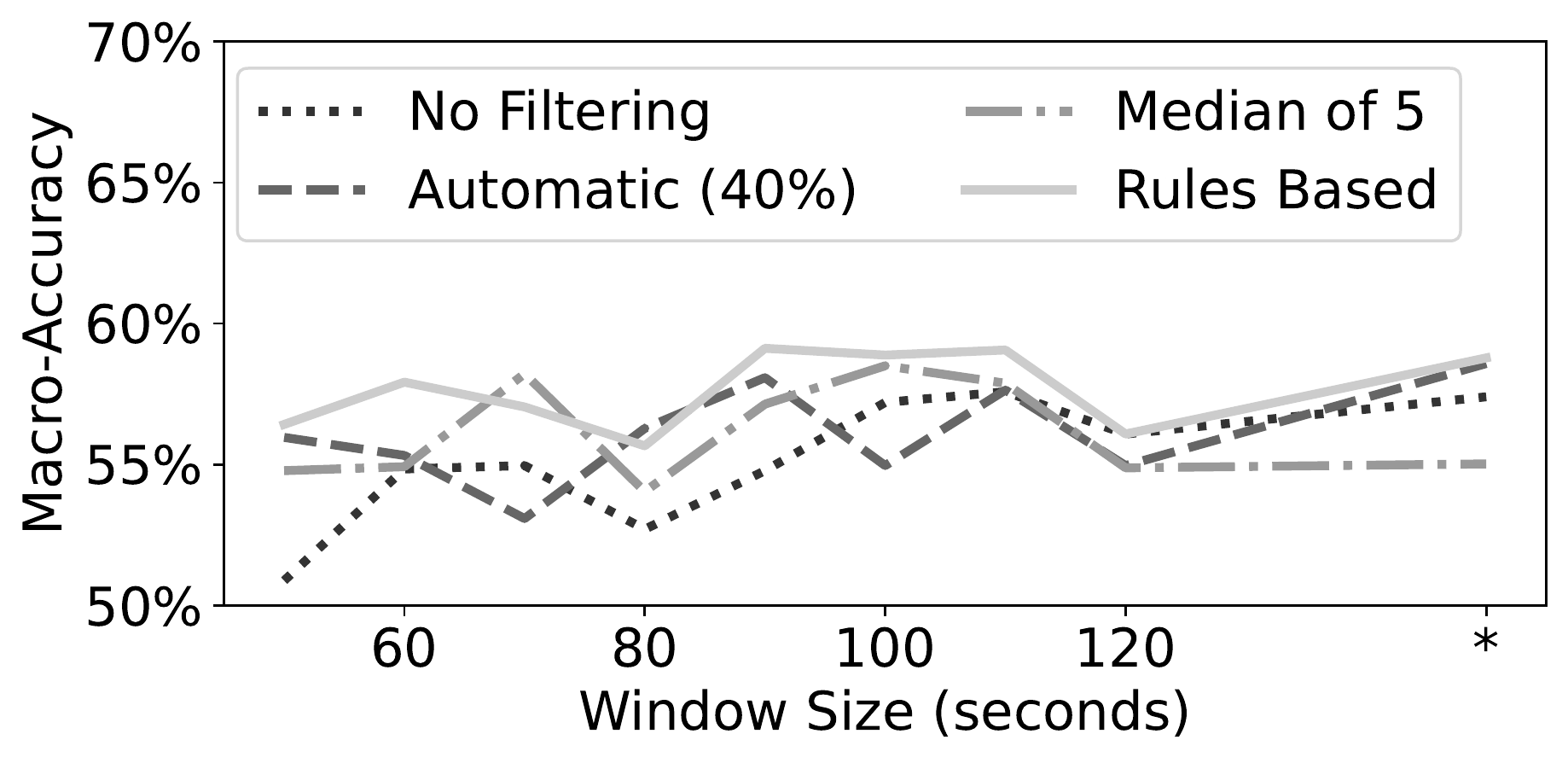}
}
\end{minipage}
\hfill
\begin{minipage}[t]{0.48\linewidth}
\vskip 0pt
{
    \caption{There is no ``best'' NN interval filter with respect to RF model performance on the five tasks in Section~\ref{sec:results}. We only include NN features in this analysis and follow an NLOPOCV procedure to evaluate model performance. The asterisk indicates using all data (~3 minutes). The automatic filter is in \cite{karlsson2012automatic}. The rules filter dropped intervals outside of a threshold.} 
    \label{fig:filter}
}
\end{minipage}
\vspace{-.5cm}
\end{figure*}

\paragraph{NN Intervals} We evaluated several methods to remove noise from the raw NN intervals. Using nested leave one participant out cross-validation (NLOPOCV) we trained random forest (RF) models to predict each of the tasks described in Section~\ref{sec:results} from NN features (c.f., Table~\ref{tab:feats}). We considered an NN interval cleaning method superior if its downstream RF models had higher macro-averaged accuracy (macro-accuracy) averaged over all prediction tasks. We did not find any one method to be universally ``best'' (see Figure~\ref{fig:filter}). For smaller time windows, the median filter appears best, while longer time windows appear to benefit from the automatic filter \cite{karlsson2012automatic} (with a 40\% cut-off rule) .

\paragraph{EDA} We applied a lowpass $4^{th}$ order Butterworth filter with a high cut of 3Hz to the raw EDA signal. As was the case with PPG, the cleaning of EDA was applied using Neurokit2 \cite{Makowski2021neurokit}.

\paragraph{Accelerometer and Skin Temperature} No filtering or cleaning was applied to accelerometer and skin temperature readings. This is appropriate for the features derived from these sensors (i.e., mean and standard deviation) under the assumption that noise on these sensors is additive and zero-mean.

\subsection{Features}

To answer our questions, we select thirteen features from the literature that have been found to accurately predict anxiety and/or stress and that can be calculated from our sensor streams. These features are calculated, from the cleaned data, for each event-phase interval per participant. Calculating the features gives a dataset with $594$ samples of the $13$ features. The full list of features can be seen in Table~\ref{tab:feats}. 

Ten of the features belong to the time-domain. We calculate these directly from the cleaned data. Two of the features are statistics of the frequency-domain, requiring a transformation from the time-domain to the frequency-domain. For this transformation we apply the Lomb-Scargle method to the time-domain samples using evenly spaced basis functions between $0.01$ and $0.5$Hz.

\subsection{Feature Correlations}

\begin{figure}[!t]
\begin{minipage}[t]{.48\linewidth}
\vskip 0pt
{
    \begin{center}
        \includegraphics[width=.95\textwidth]{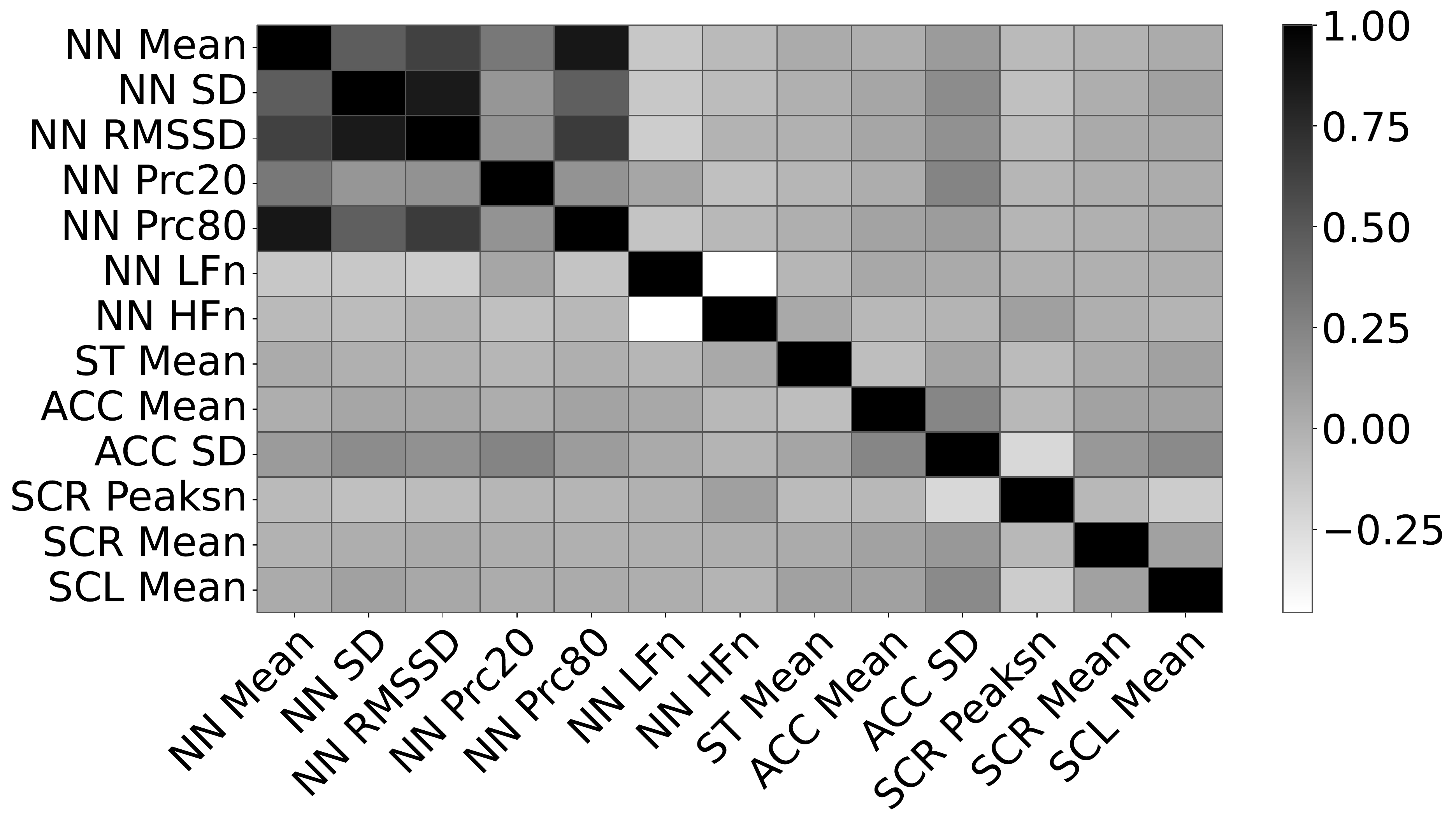}
        \captionof{figure}{The feature correlation matrix for the complete dataset. The majority of features have low correlations, suggesting each feature provides distinct information. \label{fig:corr}}
    \end{center}
}
\end{minipage}
\hfill
\begin{minipage}[t]{0.48\linewidth}
\vskip 0pt
{
    \begin{center}
    \captionof{table}{We achieve the best mean macro-accuracy across all tasks examined in Section \ref{sec:results} when we pre-condition by centering features at the participant level but don't scale. The reported performance is realized by RF models trained via NLOPOCV.}
    \vspace{.51cm}
    \begin{tabular}{ c c c c }
    \toprule
      \multicolumn{2}{c}{Participant} &
      \multicolumn{2}{c}{RF Macro-Accuracy} \\
          \cmidrule(r){1-2} 
            \cmidrule(r){3-4}
        Center & Scale & Mean & SEM \\
    \midrule
        $\checkmark$ & $\checkmark$  & $61.3\%$          & $2.3\%$ \\
        $\checkmark$ & X             & $\mathbf{65.7\%}$ & $2.4\%$ \\
         X            & $\checkmark$ & $61.7\%$          & $2.1\%$ \\
         X            & X            & $61.3\%$          & $2.1\%$ \\
    \bottomrule
    \end{tabular}
    \label{tab:cond}
    \end{center}
}
\end{minipage}
\end{figure}

A correlation matrix for the 13 features can be seen in Figure~\ref{fig:corr}. The majority of features have small inter-correlations, which suggests that each feature is providing distinct information. The features exhibiting the highest correlations are all derived from the PPG sensor.

\subsection{Feature Conditioning}

Before conducting the analysis, we considered whether to center or scale data at the participant level. To center, we subtract a participant's median feature value feature-wise. To scale, participant features are divided by the participant's feature interquartile range. See Table-\ref{tab:cond} for the mean RF model macro-accuracy across all tasks in Section~\ref{sec:results} under different conditions. Based on the results, we center but do not scale at the participant level before performing the RF analysis in Results. It is worth noting that, while scaling is typically not needed for RF models, in this case it is appropriate because it is being applied to each participant separately.

\begin{table*}
    \centering
    \caption{The thirteen features used in this study. }{
    \begin{tabular}{ l l l l l }
    \toprule
        Feature & Sensor & Units & Citation & Description\\
    \midrule
    NN Mean  & PPG & Seconds              & \cite{sun2012activity,hickey2021smart} & Mean of NN intervals \\ 
    NN SD    & PPG & Seconds              & \cite{hickey2021smart} & Standard deviation of NN intervals\\
    NN RMSSD  & PPG & Seconds             & \cite{barreda2020users,hickey2021smart,hernando2018validation} & Root mean square of successive diffs\\
    NN Prc20 & PPG & Seconds              & \cite{Makowski2021neurokit} & 20th percentile of NN intervals\\
    NN Prc80 & PPG & Seconds              & \cite{plarre2011continuous} & 80th percentile of NN intervals\\
    NN LFn   & PPG & \% Power             & \cite{hickey2021smart,hernando2018validation,ahn2019novel} & Power in 0.04-0.15 Hz / total power  \\
    NN HFn   & PPG & \% Power             & \cite{hickey2021smart,hernando2018validation,ahn2019novel} & Power in 0.15-0.40 Hz / total power \\
    ST Mean  & TMP & $^{\circ}$Celsius    & \cite{hickey2021smart} & Mean of temperature samples \\
    ACC Mean & ACC & G-force              & \cite{boukhechba2018demonicsalmon} & Mean of $||[x,y,z]||$ \\
    ACC Sd   & ACC & G-force              & \cite{boukhechba2018demonicsalmon} & Standard deviation of $||[x,y,z]||$ \\
    SCL Mean & EDA & \SI{}{\micro Siemens}& \cite{barreda2020users,petrescu2020integrating} & Mean of tonic amplitudes \\
    SCR Mean & EDA & \SI{}{\micro Siemens}& \cite{barreda2020users,petrescu2020integrating} & Mean of phasic amplitudes \\
    SCR Peaksn & EDA & Peaks/Time         & \cite{sun2012activity,ihmig2020line} & Count of SCR peaks / total time. \\
    \bottomrule
    \end{tabular}}
    \label{tab:feats}
\end{table*}

\section{Results}
\label{sec:results}

We answer four questions regarding participants in our study:
\begin{enumerate}
    \item How do biomarkers differ between being alone vs. in a social situation?
    \item How do biomarkers differ between the pre-event, event, and post-event phases of social situations? 
    \item How do biomarkers differ between dyadic vs. group interactions?
    \item How do biomarkers differ between implicit and explicit evaluative interactions?
\end{enumerate}

For the analyses that follow, statistical significance is determined via a Wilcoxon signed-rank test and the Benjamini-Hochberg procedure at the task level, with $\alpha=.05$. Plotted point estimates for features represent the medians with a 95\% boostrap CI (both of which more naturally align with the non-parametric Wilcoxon signed-rank test). Plots that compare model performance present the average of performance across folds in NLOPOCV. The performance metric, macro-accuracy, is macro-averaged accuracy to control for class imbalance in testing data. In all of our analysis baseline performance for the macro-average is 50\%. Error bars in model performance figures represent the standard error. Hyperparameter tuning was applied in the inner cross-validation loop to determine the RF complexity pruning as well as the feature selection method (i.e., a mutual information criteria or ANOVA F-value).

\subsection{Social Context}

To compare passive features across social context (i.e., alone vs. social tasks), we separate the event observations into an alone set, containing only data from the concurrent phase of the alone task, and a social set, containing data from the concurrent phase of all dyadic and group conversations.

\begin{figure}
    \begin{minipage}{0.29\textwidth}
        \begin{subfigure}{\textwidth}
            \includegraphics[width=\textwidth]{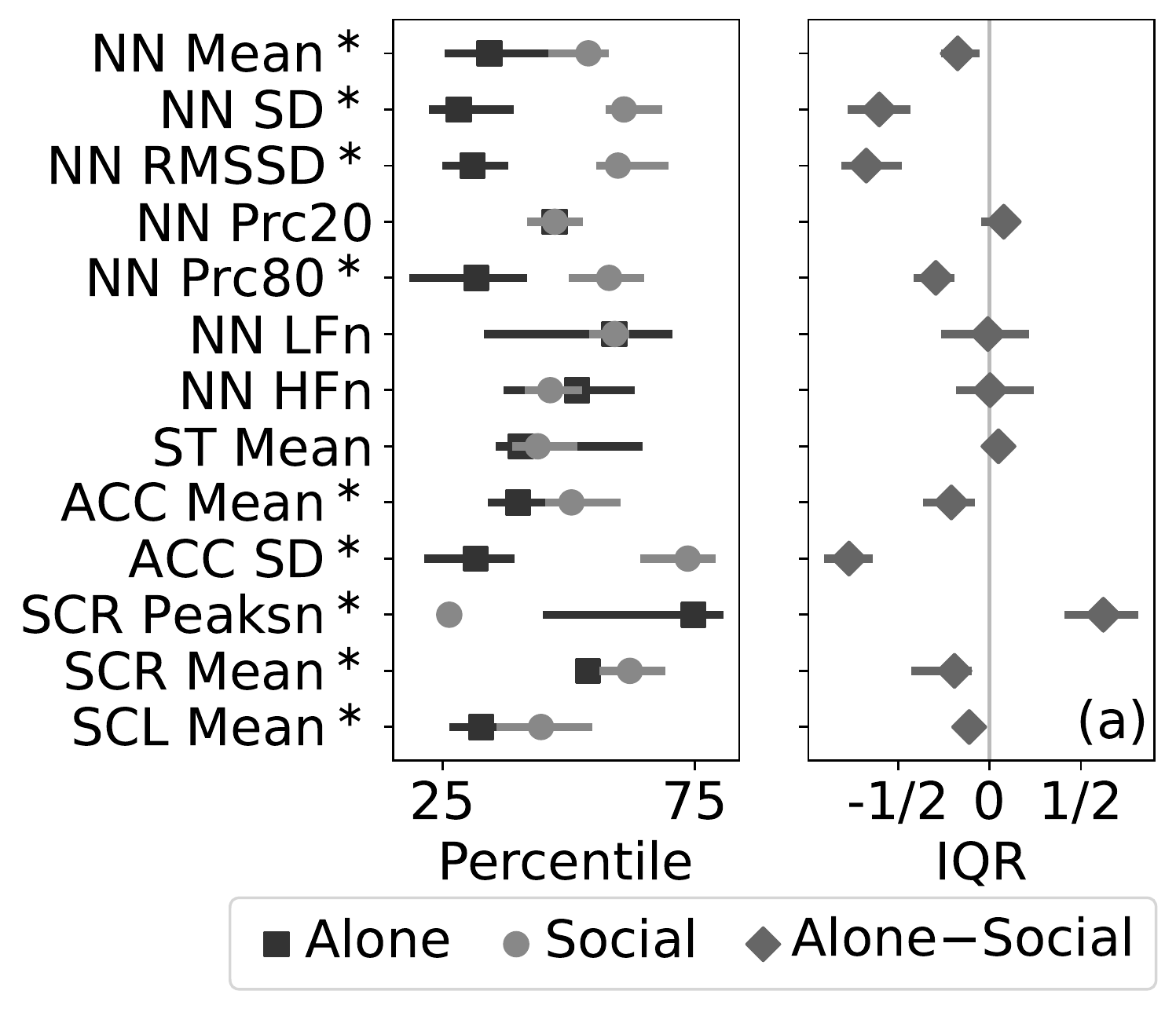}
        \end{subfigure}
    \end{minipage}    
    \hfill
    \begin{minipage}{0.25\textwidth}
        \begin{subfigure}{\textwidth}
        \centering
        \includegraphics[width=.93\textwidth]{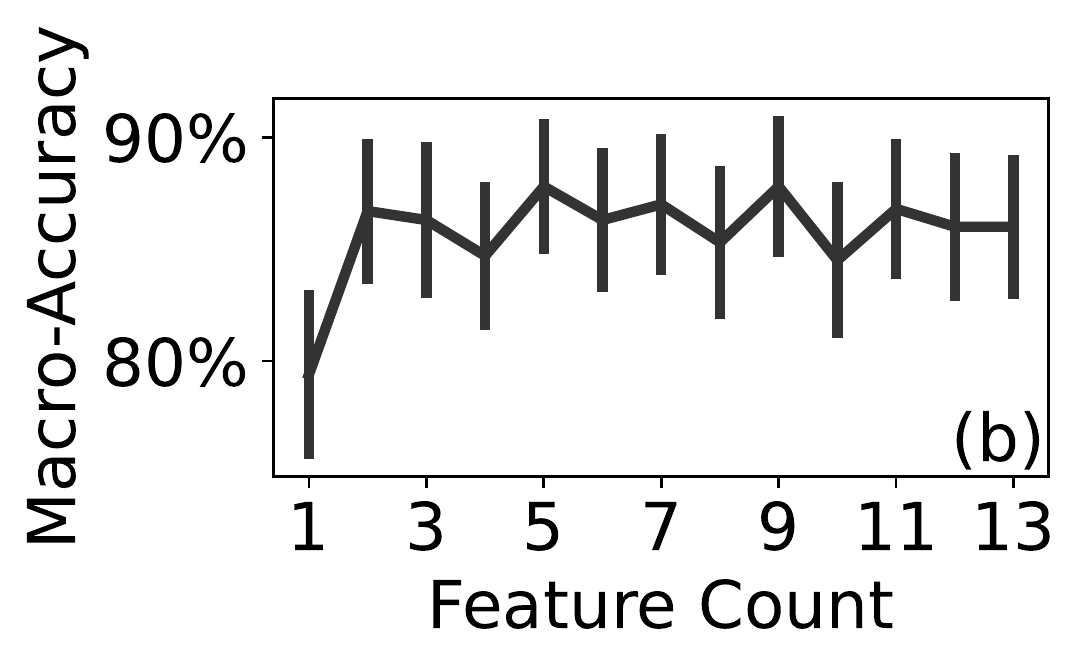}
        \end{subfigure}
        \vfill
        \begin{subfigure}{\textwidth}
        \centering
        \includegraphics[width=.96\textwidth]{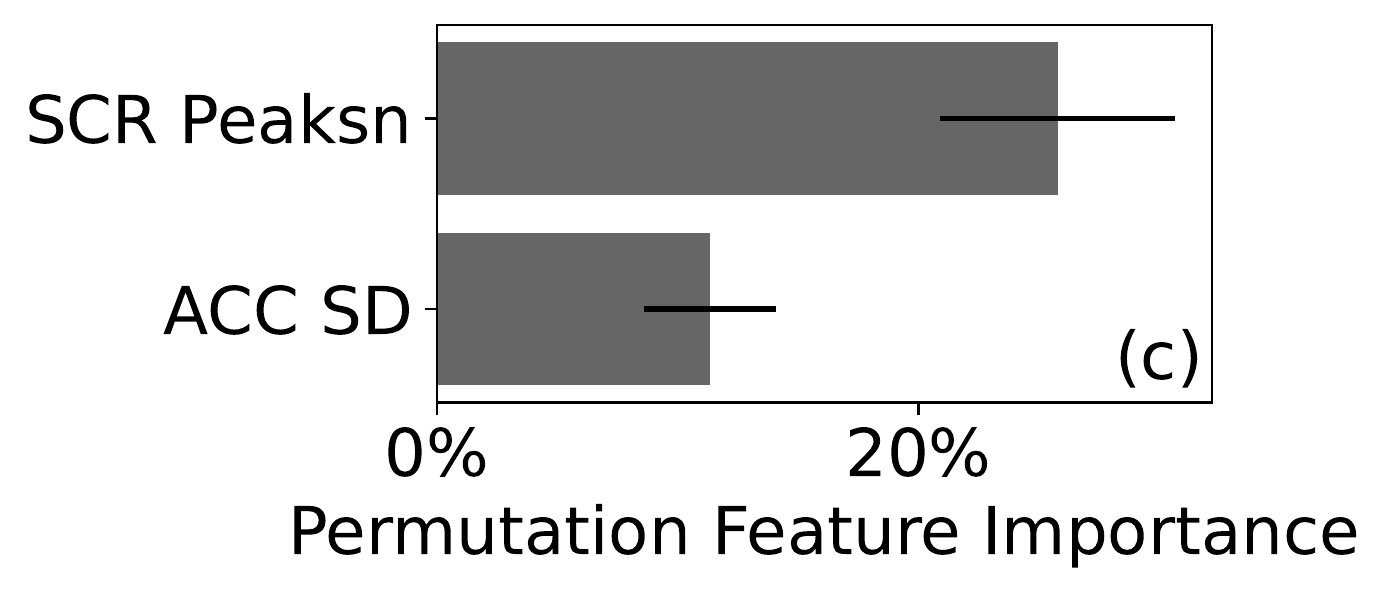}
        \end{subfigure}
    \end{minipage}
    \hfill
    \begin{minipage}{0.39\textwidth}
        \caption{(a) Nine features have significant paired differences between the alone and social tasks. Plotted are medians with a 95\% boostrap CI. (b) Using NLOPOCV we estimate an RF model achieves near peak performance with two features. (c) We show feature importance for the two feature model by calculating the decrease in macro-accuracy when a feature is permuted.}
        \label{fig:alone_social_feats}
        \label{fig:alone_social_import}
        \label{fig:alone_social_kbest}
    \end{minipage}
\end{figure}

There are nine features that are significantly different in the alone vs. social set (see Figure~\ref{fig:alone_social_feats}~(a)).  The three largest differences are NN RMSSD, ACC SD, and SCR Peaksn. We perform a multivariate analysis to evaluate the independence of each feature. Using NLOPOCV we determine that an RF model predicting whether a test example belongs to the alone or social set can achieve peak performance using only two features (see Figure~\ref{fig:alone_social_kbest}~(b)).

Using the full dataset to select the two best features, we fit one more random forest to examine the importance of each feature via permutation (see Figure~\ref{fig:alone_social_import}~(c)). Following this procedure produces a model which only uses features from the EDA and ACC sensors. We can infer from Figure~\ref{fig:alone_social_kbest}~(b) that additional features from the PPG and TMP sensors do not provide any additional information to the model. This is in spite of the fact that several PPG derived features had significantly different medians between the two sets.

\subsection{Social Timing}
To understand how features change with respect to timing, we perform two analyses: (1) compare features from social events to their combined pre and post phases, and (2) compare features recorded during the pre-social task phase to the post-social task phase across all social tasks. Our comparison of the pre- and post-event social phases was motivated by literature suggesting that these phases may be characterized by related but distinct processes. For example, anticipatory (or pre-event) anxiety may resemble in-the-moment anxiety more closely and involve physiological arousal whereas post-event processing may be a primarily cognitive process \cite{heimberg_cognitive-behavioral_2014}.

\subsubsection{Event vs. Pre/Post}

\begin{figure}
    \begin{minipage}{0.38\textwidth}
        \caption{ (a) Twelve features have significant paired differences between social events and their pre/post phases. Plotted are medians with a 95\% boostrap CI. (b) Using NLOPOCV we estimate an RF model realizes peak performance with thirteen features. (c) We show feature importance for the five feature model via the decrease in macro-accuracy when we permute a feature.}
        \label{fig:event_prepost_feats}
        \label{fig:event_prepost_import}
        \label{fig:event_prepost_kbest}
    \end{minipage}
    \begin{minipage}{0.29\textwidth}
        \begin{subfigure}{\textwidth}
            \hspace{.5cm}
            \includegraphics[width=\textwidth]{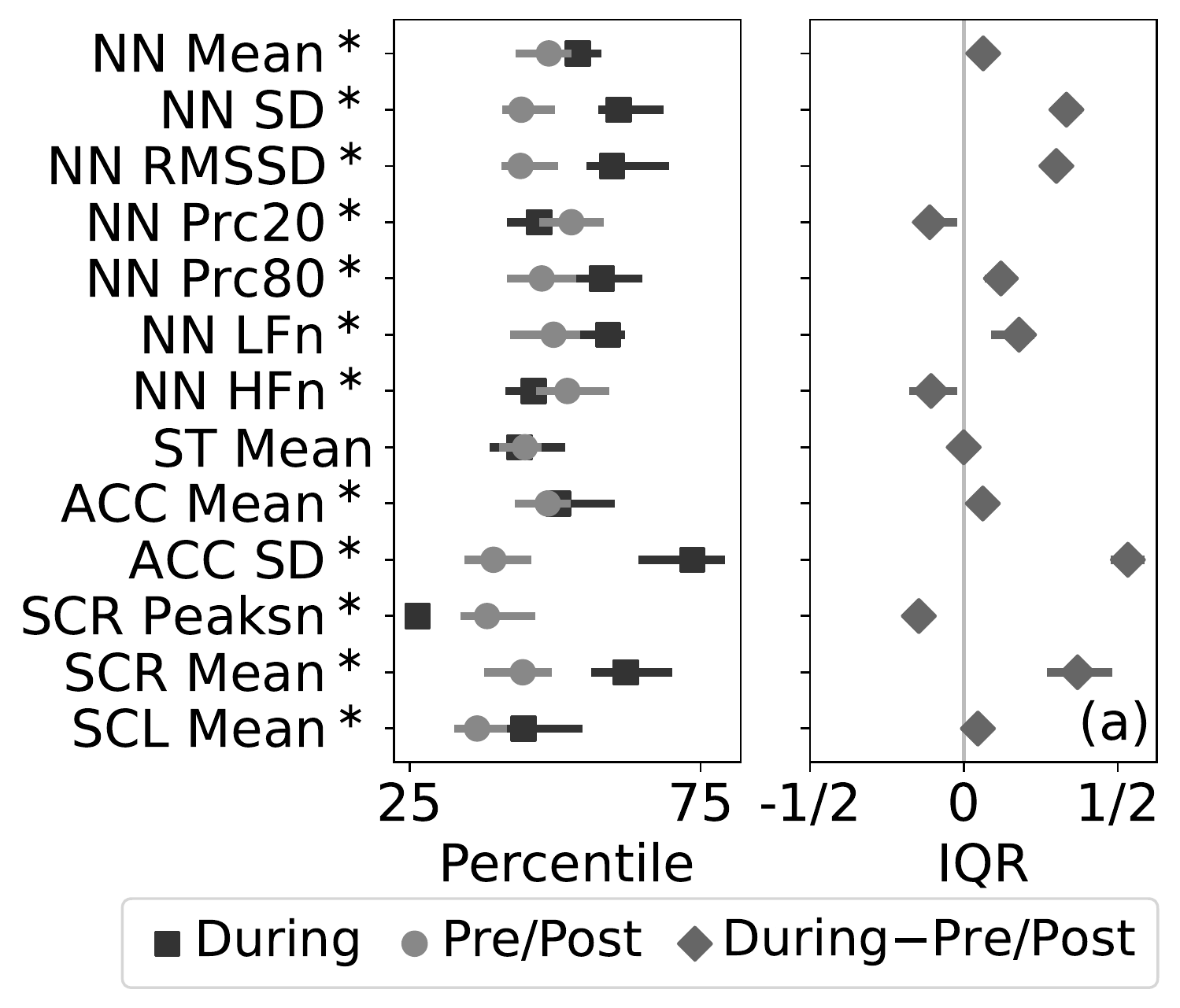}
        \end{subfigure}
    \end{minipage}    
    \hfill
    \begin{minipage}{0.25\textwidth}
        \begin{subfigure}{\textwidth}
        \centering
        \includegraphics[width=\textwidth]{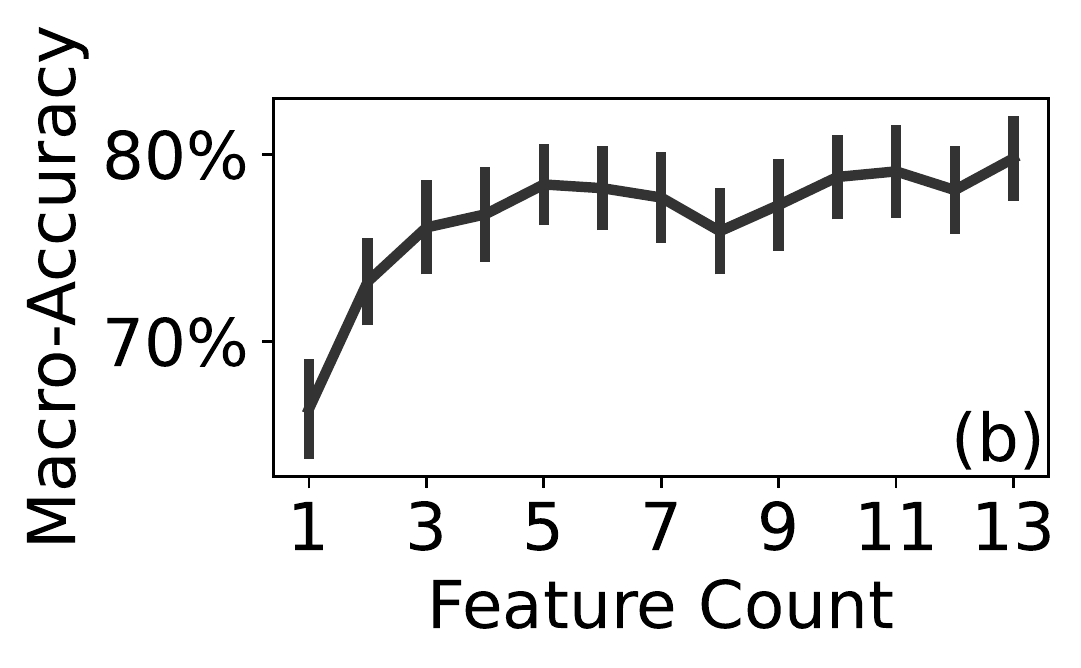}
        \end{subfigure}
        \vfill
        \begin{subfigure}{\textwidth}
        \centering
        \includegraphics[width=\textwidth]{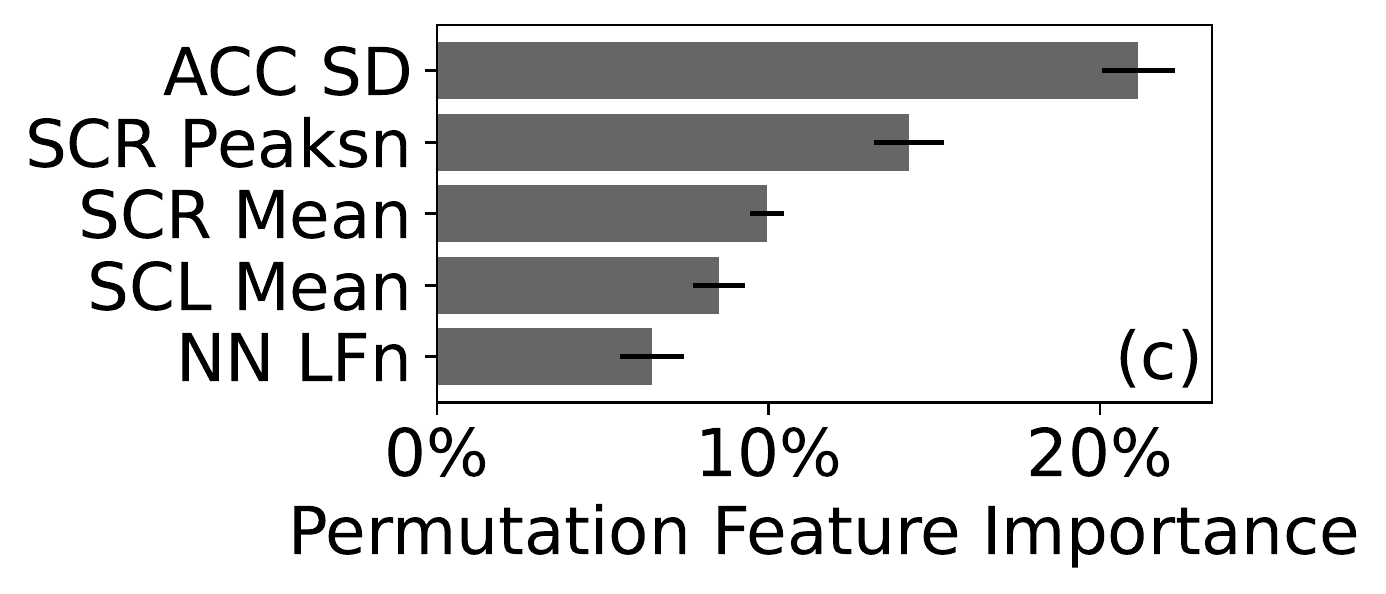}
        \end{subfigure}
    \end{minipage}
    \hfill
\end{figure}

There are twelve features with significant paired differences between a social event and its pre/post phases. This analysis excludes all data from the alone experience. The three largest significant differences are ACC SD, SCR Mean, and NN SD. The full feature analysis can be seen in Figure~\ref{fig:event_prepost_feats}~(a).

As before, we perform a multivariate analysis by fitting an RF with a small subset of features. Using NLOPOCV we determine only five features are needed for performance within one standard error of peak performance. We select the best five features using the full dataset and then use feature permutation with an RF model to determine feature importance. The order of importance is ACC SD, SCR Peaksn, SCR Mean, SCL Mean, and NN LFn (see Figure~\ref{fig:event_prepost_import}). Interestingly, while we downselect to five features for our analysis, the model appears to continue to improve all the way up to the full thirteen features.

\subsubsection{Pre vs. Post}

\begin{figure}
    \vspace{-.3cm}
    \begin{minipage}{0.29\textwidth}
        \begin{subfigure}{\textwidth}
            \includegraphics[width=\textwidth]{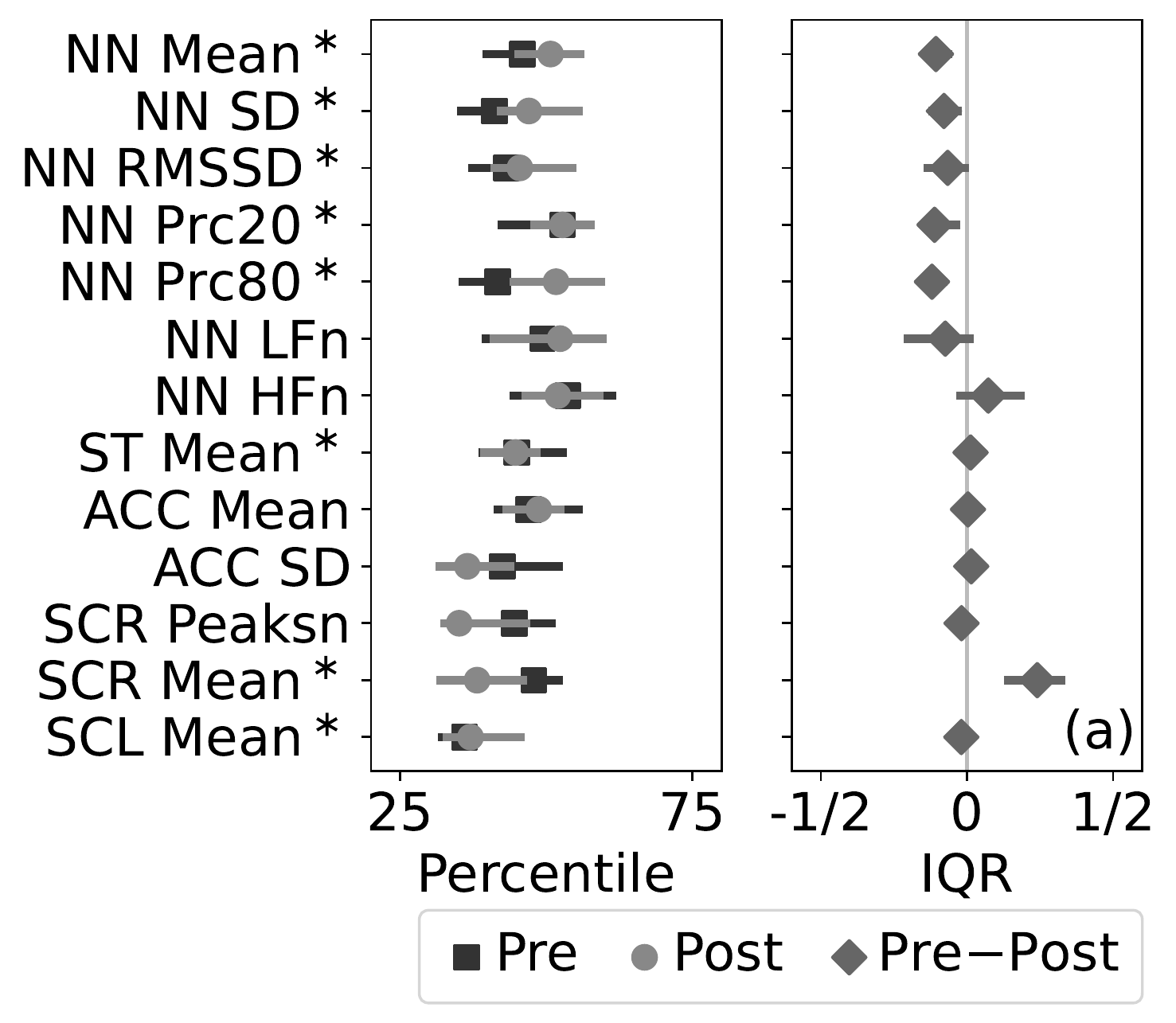}
        \end{subfigure}
    \end{minipage}    
    \hfill
    \begin{minipage}{0.25\textwidth}
        \begin{subfigure}{\textwidth}
        \centering
        \includegraphics[width=\textwidth]{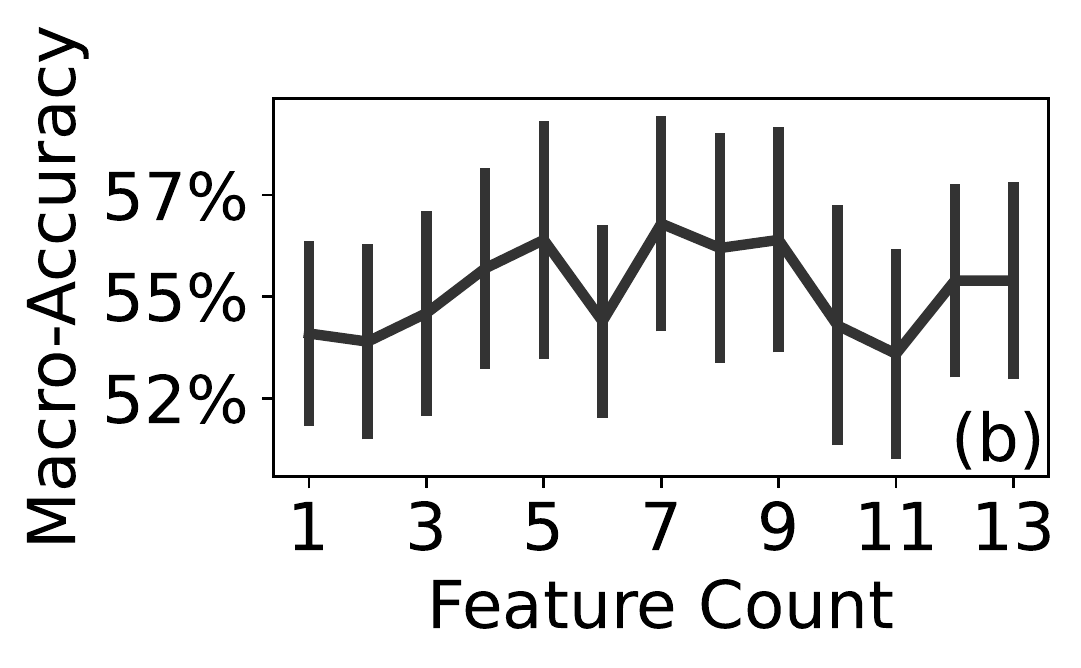}
        \end{subfigure}
        \vfill
        \begin{subfigure}{\textwidth}
        \centering
        \includegraphics[width=\textwidth]{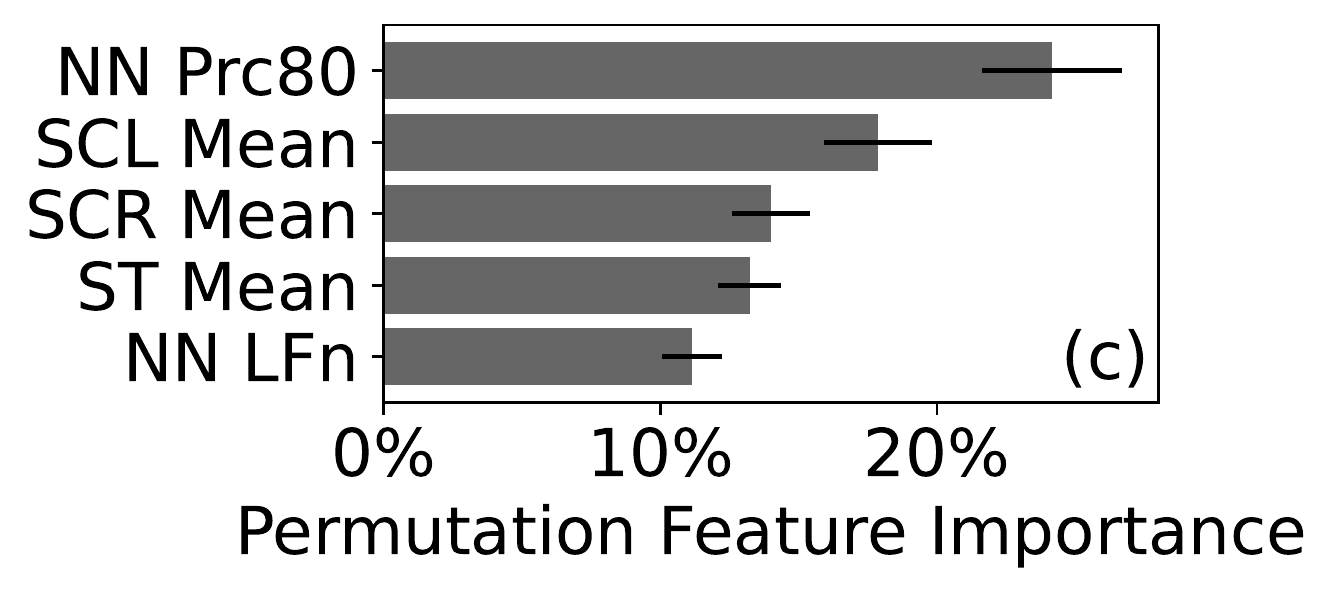}
        \end{subfigure}
    \end{minipage}
    \hfill
    \begin{minipage}{0.38\textwidth}
        \caption{(a) Eight features have significant paired differences between pre phases and post phases. Plotted are medians with a 95\% boostrap CI. (b) Using NLOPOCV we estimate that an RF model achieves peak performance with five features. (c) We show feature importance for the five feature model by calculating the decrease in macro-accuracy when a feature is permuted.}
        \label{fig:pre_post_feats}
        \label{fig:pre_post_import}
        \label{fig:pre_post_kbest}
    \end{minipage}
\end{figure}

We perform a second timing analysis comparing biomarker features between the pre-event phase and post-event phase of social tasks. This gives eight significant paired differences: five significant NN features, ST Mean, SCR Mean, and SCL Mean. The full result can be seen in Figure~\ref{fig:pre_post_feats}~(a). A multivariate analysis, using NLOPOCV to determine a minimal model, was not able to meaningfully remove any features (or even perform that well overall given all features). This shows that, given our features, it is not feasible to discriminate between pre-event and post-event phases of social situations.

\subsection{Social Interaction Size}

Using dyadic event features and group event features we repeat our analysis to determine whether the passive sensors can distinguish the number of individuals in a social interaction.
We observe only four significant feature differences: ACC SD, ACC Mean, NN Prc20, and SCR Peaksn. Fitting a random forest model we observe predictive accuracy of 58\%, which is only slightly above random guessing (i.e., 50\%), and the feature importance of the five feature model mostly relies on ACC SD.

\begin{figure}
    \begin{minipage}{0.39\textwidth}
        \caption{(a) Four features have significant paired differences between dyadic and group events. Plotted are medians with a 95\% boostrap CI. (b) Using NLOPOCV we estimate an RF model realizes peak performance with seven features. (c) We show feature importance for the five feature model by calculating the decrease in macro-accuracy when a feature is permuted.}
        \label{fig:dyadic_group_feats}
        \label{fig:dyadic_group_import}
        \label{fig:dyadic_group_kbest}
    \end{minipage}
    \hfill
    \begin{minipage}{0.29\textwidth}
        \begin{subfigure}{\textwidth}
            \includegraphics[width=\textwidth]{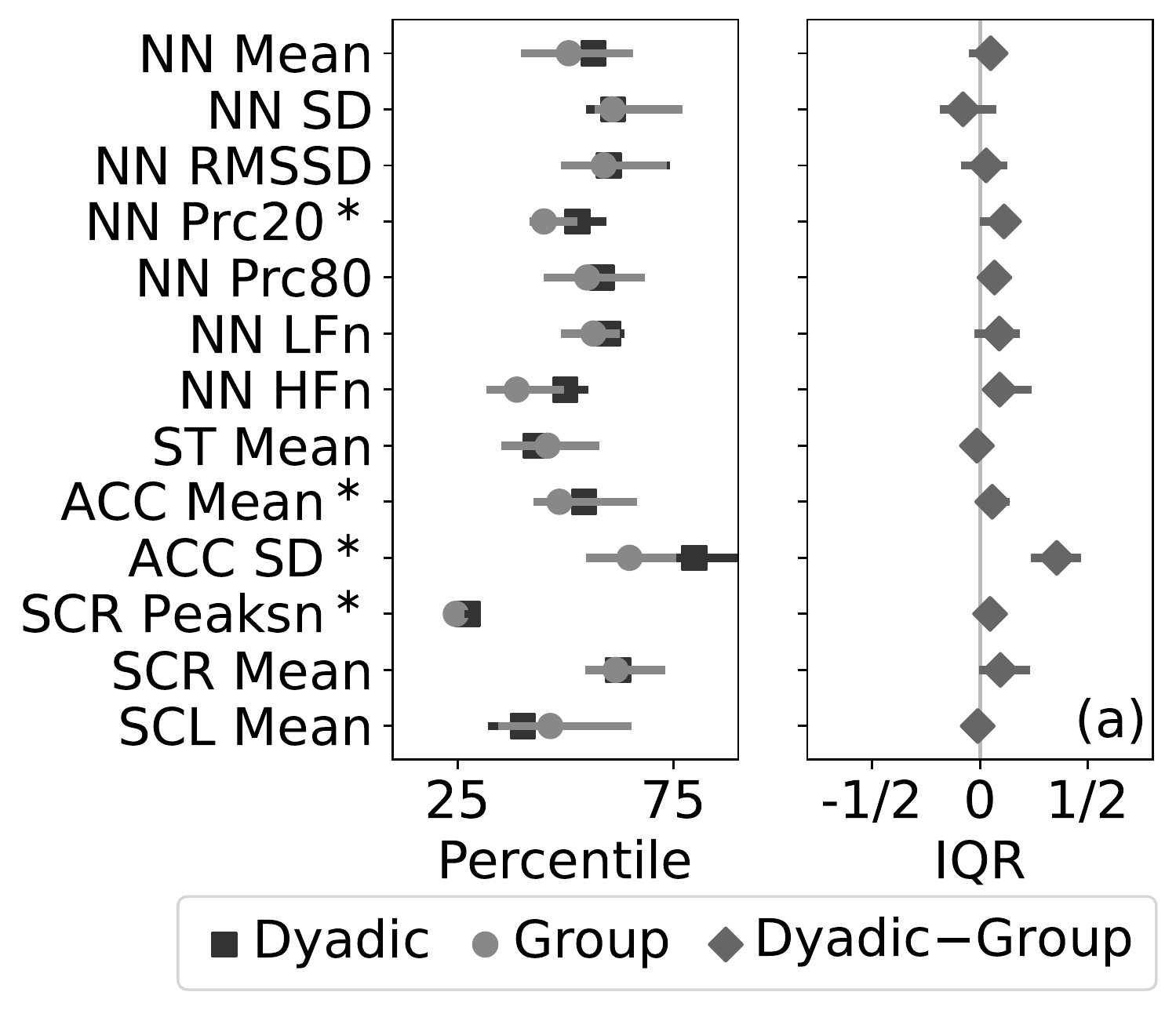}
        \end{subfigure}
    \end{minipage}    
    \hfill
    \begin{minipage}{0.25\textwidth}
        \begin{subfigure}{\textwidth}
        \centering
        \includegraphics[width=\textwidth]{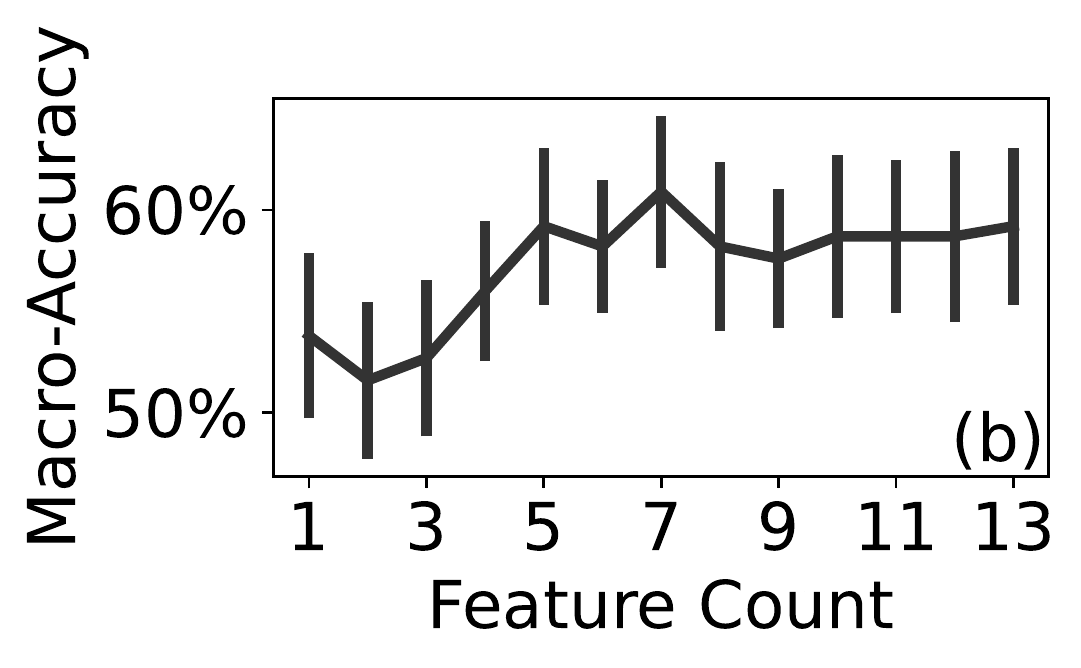}
        \end{subfigure}
        \vfill
        \begin{subfigure}{\textwidth}
        \centering
        \includegraphics[width=\textwidth]{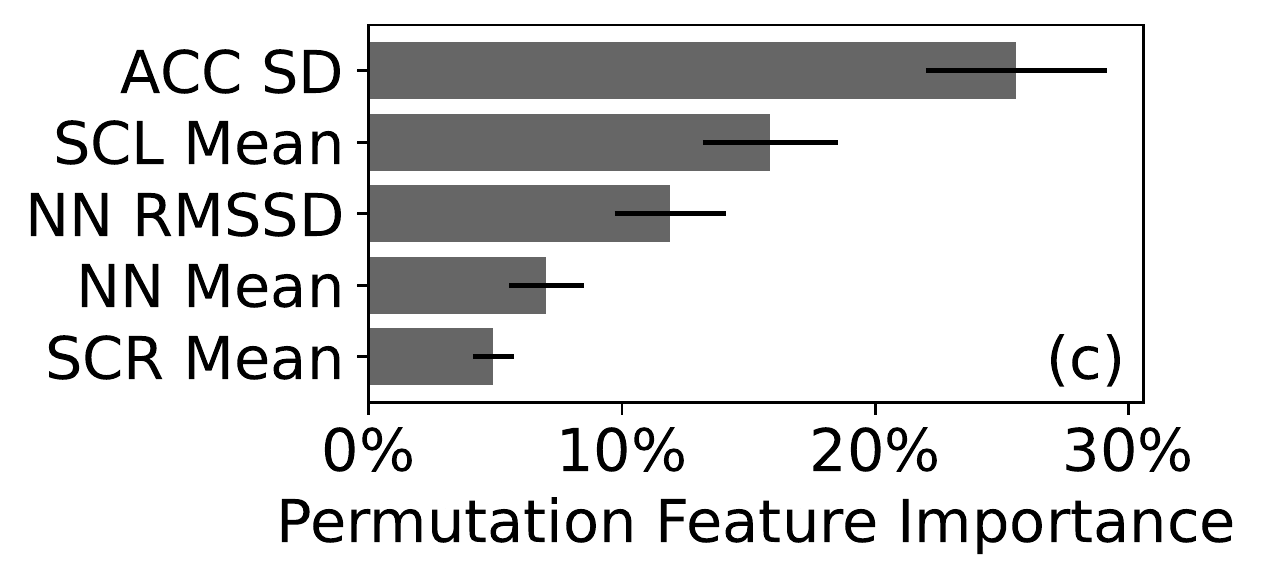}
        \end{subfigure}
    \end{minipage}
\end{figure}

\subsection{Social Threat}
Examining differences in biomarkers between implicit social threat situations and explicit social threat situations, we see only one significant feature difference, SCL Mean. Similarly, the NLOPOCV trained RF models struggle to perform above random guessing (i.e., 50\%). The fact that our models consistently perform worse than random guessing suggests that participants differ from one another in systematic ways (since NLOPOCV tests on a single participant at a time).

\begin{figure}
    \begin{minipage}{0.29\textwidth}
        \begin{subfigure}{\textwidth}
            \includegraphics[width=\textwidth]{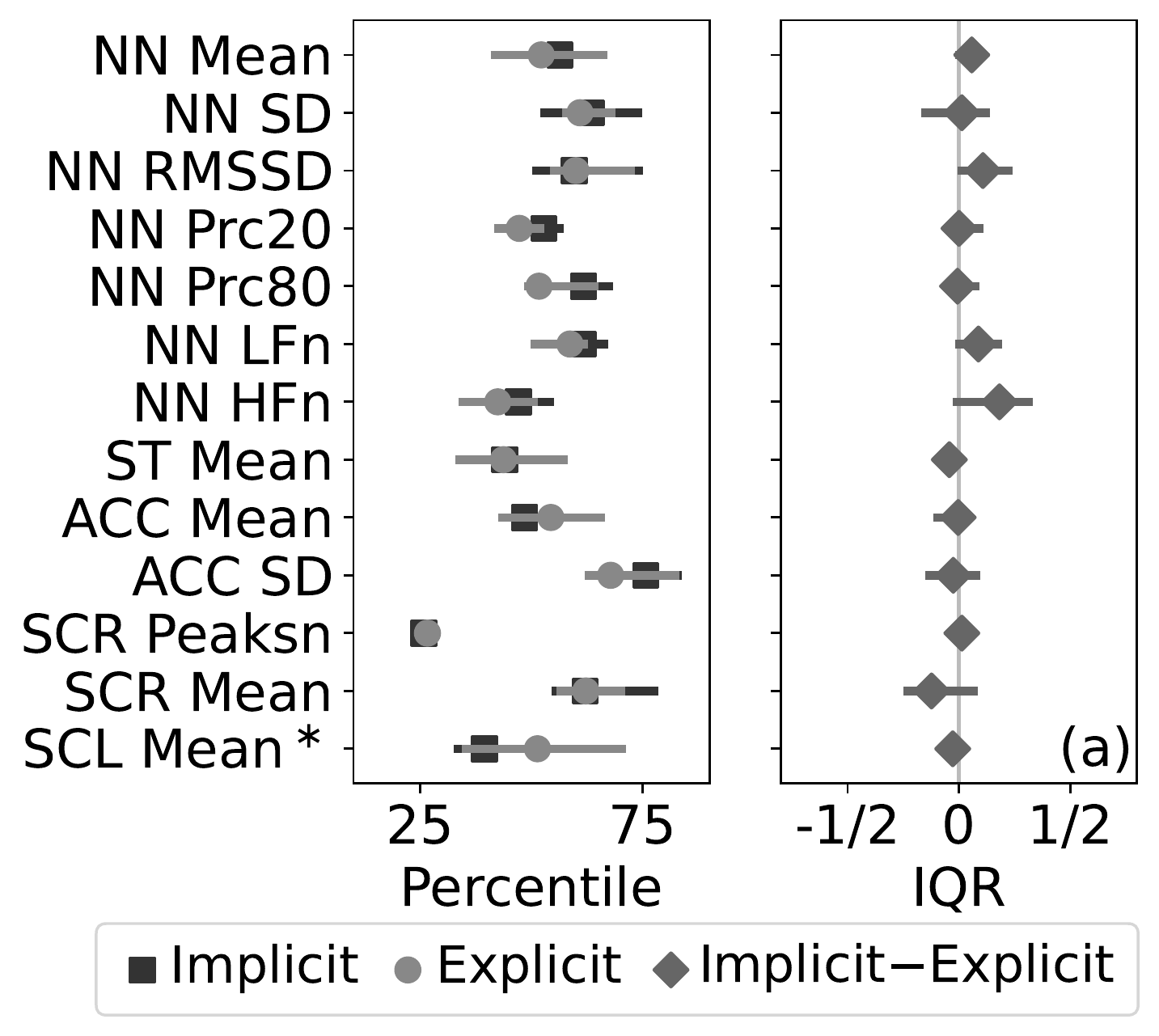}
        \end{subfigure}
    \end{minipage}    
    \hfill
    \begin{minipage}{0.25\textwidth}
        \begin{subfigure}{\textwidth}
        \centering
        \includegraphics[width=\textwidth]{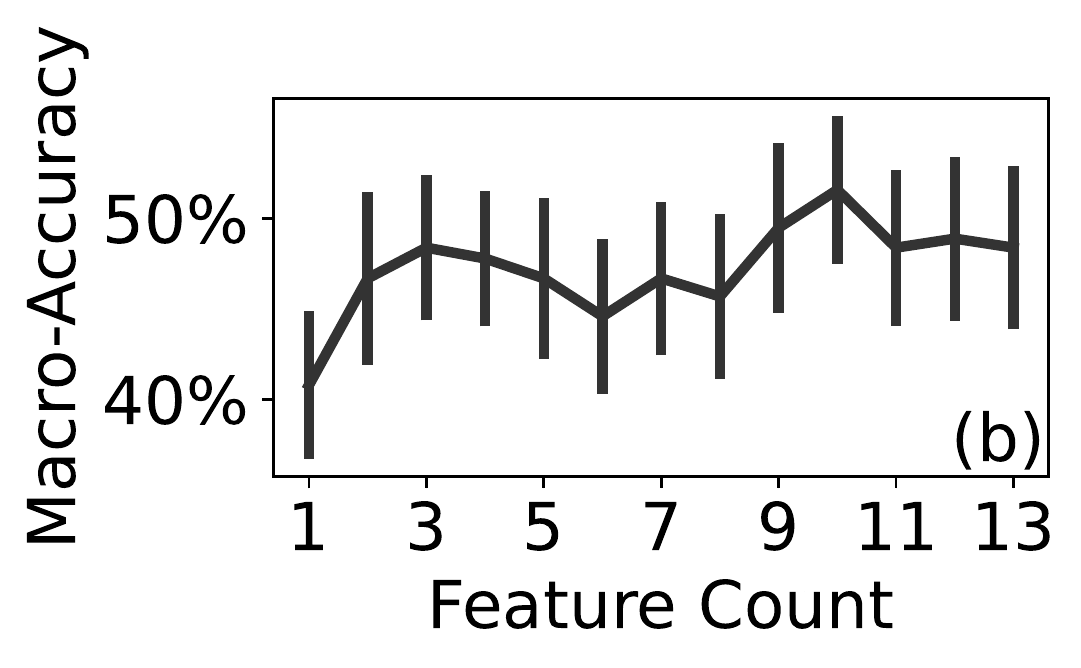}
        \end{subfigure}
        \vfill
        \begin{subfigure}{\textwidth}
        \centering
        \includegraphics[width=\textwidth]{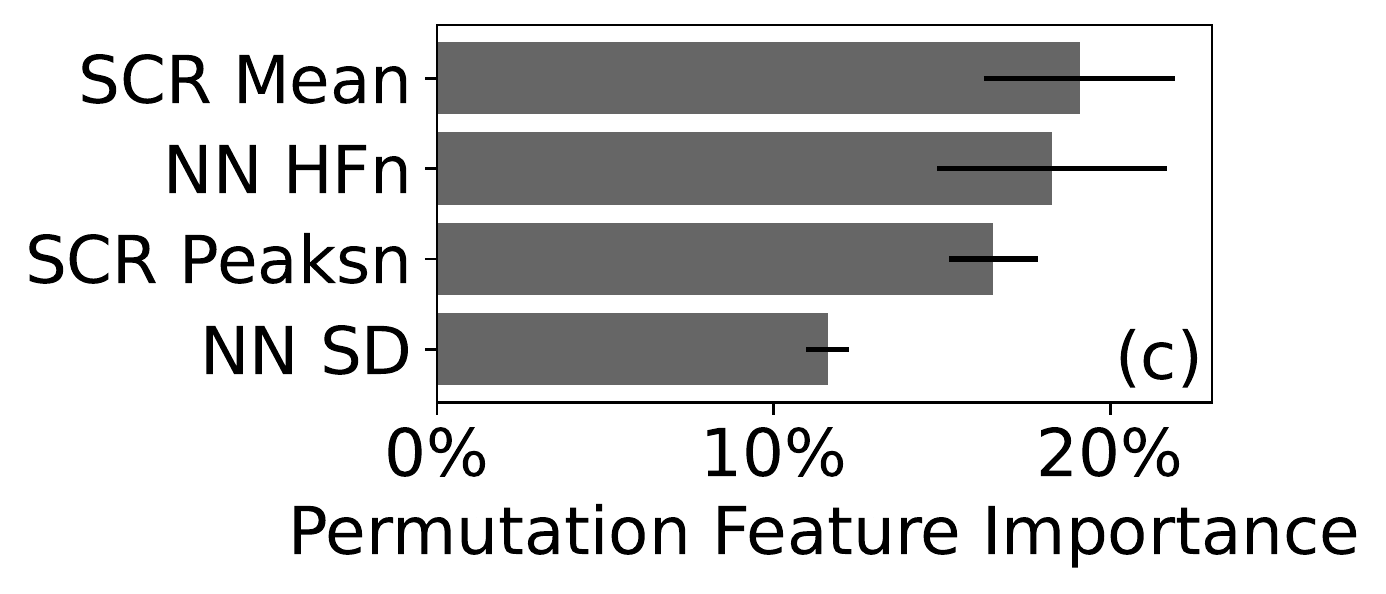}
        \end{subfigure}
    \end{minipage}
    \hfill
    \begin{minipage}{0.39\textwidth}
        \caption{(a) One feature has significant paired difference between the implicit and explicit evaluation. We plot medians with a 95\% boostrap CI. (b) Using NLOPOCV we estimate that an RF model achieves peak performance with ten features. (c) We show feature importance for the four feature model by calculating the decrease in macro-accuracy when a feature is permuted.}
        \label{fig:implicit_explicit_feats}
        \label{fig:implicit_explicit_import}
        \label{fig:implicit_explicit_kbest}
    \end{minipage}
\end{figure}

\subsection{Post-hoc Analysis}

Existing work \cite{siegel2018emotion,hoemann2020context} reveals that physiological data can exhibit considerable heterogeneity across individuals. Motivated by this we perform a post-hoc cluster analysis to evaluate the heterogeneity of our own data. For this analysis we focus on two tasks presented in Section~\ref{sec:results}: (1) alone vs. social, and (2) during/concurrent vs. pre/post. We select these tasks because they are the only two tasks whose predictive models significantly outperformed the random guessing baseline of 50\%. Before clustering we select down to the most important features for prediction in each task (see Figure~\ref{fig:alone_social_feats}~(c) and Figure~\ref{fig:event_prepost_feats}~(c)). We then apply Hierarchical Density-Based Spatial Clustering of Applications with Noise (HDBSCAN) \cite{McInnes2017} to our data. This algorithm both determines the appropriate number of clusters and filters out outlier data points.

\begin{figure}[t]
    \centering
    \begin{minipage}{0.5\textwidth}
        \includegraphics[width=\textwidth]{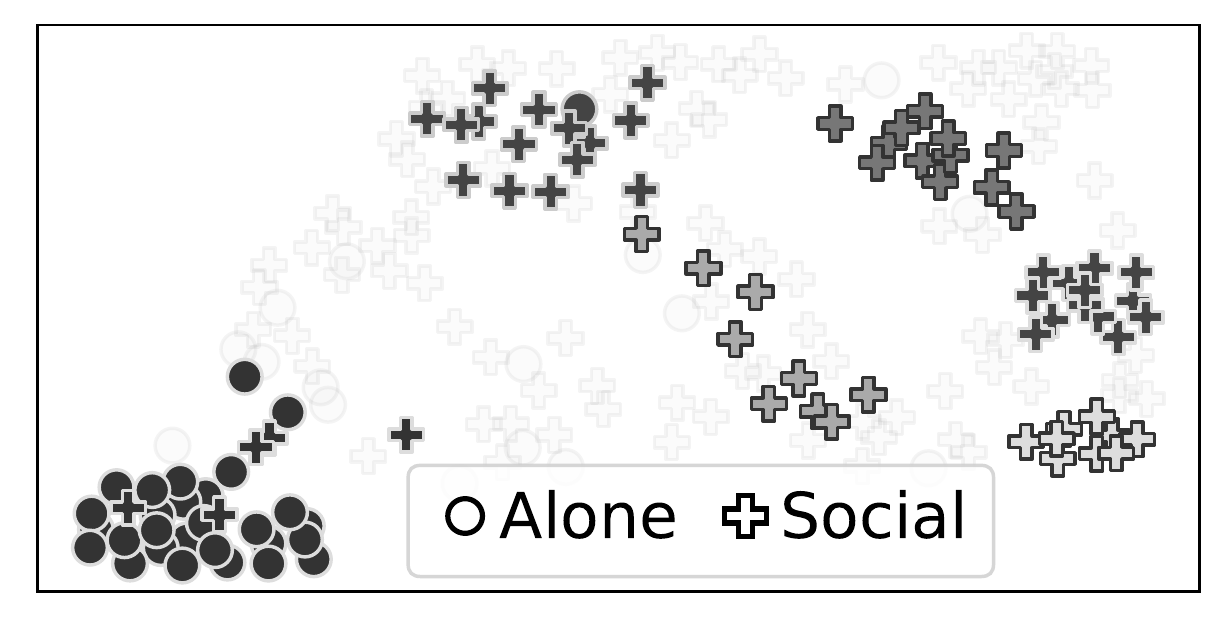}    
    \end{minipage}
    \hfill
    \begin{minipage}{0.46\textwidth}
        \caption{ Samples from the social experiences create six distinct clusters while non-social samples largely belong to a single cluster. This two-dimensional plot is a TSNE representation of the true underlying data. This is why some of the clusters may seem more spread out than one might expect. }
        \label{fig:cluster1}
    \end{minipage}
    \vspace{-.5cm}
\end{figure}

For alone vs. social, HDBSCAN finds six clusters with an average purity over 90\% (purity is the percentage of points in the cluster belonging to the dominant set). Additionally, 54\% of points are marked as outliers and therefore not assigned to any cluster. For concurrent vs. pre/post, HDBSCAN finds 22 clusters. The clusters that mostly contain data collected in the ``during" events the tasks have an average purity of 74\% while the clusters containing mostly pre/post data have an average purity of 85\%. The full results of this analysis can be seen in Table~\ref{tab:cluster} and a visualization of alone vs. social clustering can be seen in Figure~\ref{fig:cluster1}. 

\begin{table}
    \centering
    \caption{Across the first two tasks we identify 28 underlying clusters. This suggests a high-amount of heterogeneity within the data. The one exception to this is the alone dataset from alone vs. social. Almost all alone example points either belong to a single cluster or are considered outliers. In the table below purity is reported as the percentage of the majority label in the cluster.}{
    \vspace{.5cm}
    \begin{tabular}{ c c c c c c }
    \toprule
      \multicolumn{2}{c}{Group} &
      \multicolumn{3}{c}{Clusters} &
      \multicolumn{1}{c}{Outliers} \\
    \cmidrule(r){1-2} 
    \cmidrule(r){3-5}
    \cmidrule(r){6-6}
      Task & Label & Count & E[Size] & E[Purity] & Count \\
    \midrule
     Alone vs. Social    & Alone  & 1 & 33   & 84.8\% & 16 \\ 
     Alone vs. Social    & Social & 5 & 11.8 & 98.8\% & 95 \\
     During vs. Pre/Post & During & 7 & 10 & 74\% &  73 \\
     During vs. Pre/Post & Pre/Post & 15 & 11.8 & 85\% &  129 \\
    \bottomrule
    \end{tabular}}
    \label{tab:cluster}
\end{table}

\section{Discussion}
The goal of the present study was to describe physiological patterns associated with different social and non-social tasks, determine if and how these patterns differ across contexts, phases, and levels of evaluation, and explore whether socially anxious individuals are characterized by common physiological response patterns across tasks. To explore these questions, we: (1) examined whether passively sensed biomarkers relevant to anxiety and stress detection differed across social vs. non-social tasks, phases of social interactions, social interaction group sizes, and levels of evaluative threat; (2) used random forest models to identify how many features are required for optimal model performance and the most important individual features for context detection; and (3) used post-hoc clustering analyses to explore whether participants were characterized by common patterns in physiological responding during social compared to non-social tasks.

Our results suggest that the passively sensed biomarkers used in this study can reliably distinguish between periods when individuals are engaged in social interaction vs. not (i.e., social interaction vs. non-social task; during social interaction vs. pre/post social interaction). However, these physiological features could not distinguish between anticipatory/pre-event anxiety and post-event processing immediately before and after a social event, nor could they distinguish well between different types of social interactions characterized by nuanced differences (e.g., dyadic vs. group interactions; implicit vs. explicit social evaluation). These findings are in line with previous research demonstrating that it is feasible to use passively sensed biomarkers to distinguish between baseline, anticipatory, and concurrent anxiety phases \cite{shaukat-jali_detecting_2021, pisanski_multimodal_2018}. However, they are inconsistent with studies that have successfully used passive sensing to differentiate between different contextual features of social interactions. For instance, one study identified differences in heart rate, skin conductance, and voice features across levels of social evaluation during a virtual reality speech task \cite{barreda-angeles_users_2020} and another used location data to identify different group sizes during social interactions \cite{zakaria_stressmon_2019}, though the latter relied on WiFi and real-time location services to identify people who were near each other and wearing the same device. In the present study, it may be that the experimental manipulations (e.g., telling people they would be evaluated vs. not) did not produce enough of a change in anxiety such that it could be detected physiologically. Indeed, it was common for anxious responses to exhibit "discordance" \cite{hollenstein_models_2014} or a lack of strong associations among components of anxiety (e.g., heightened threat perception without accompanying physiological arousal). Consistent with this, previous work using this dataset did not find evidence for concordance among affective, cognitive, behavioral, and physiological components of social anxiety during social interactions \cite{toner_social_2022}. 

Interestingly, there were some cases in which we observed significant paired differences in median values of a given feature across contexts but the predictive models were unable to reliably distinguish between these contexts. This pattern of results suggests that these features may differ across contexts in terms of their central tendency but that there is substantial overlap and variability in the data across the full time window. In other words, significant differences between features could sometimes be observed by leveraging data from multiple participants to increase the power of our tests. However, this did not necessarily translate to an effective model since the model only has access to a single example at the time of prediction. 

Regarding specific, feature-level findings, the accelerometer standard deviation consistently emerged as the feature most able to distinguish between social and non-social contexts. This is likely reflective of the fact that participants were only speaking while actively participating in one of the social interaction tasks and were otherwise asked to sit quietly at their computer. Given that people tend to gesture while speaking \cite{goldin-meadow_role_1999}, it is understandable why this feature may be particularly useful in detecting situational contexts involving conversation. These findings are also consistent with previous work demonstrating that accelerometer data can be used to accurately detect when people are speaking in a conversation \cite{hung_detecting_2014}. Additionally, accelerometer standard deviation was significantly lower during group vs. dyadic interactions and emerged as the most informative feature in the social group size model, which may be explained by patterns of conversational turn-taking \cite{stivers_universals_2009}. That is, there were likely more opportunities (and a greater expectation) to speak, and thus gesture, during dyadic conversations as compared to group conversations based on the number of people present in the conversation. Taken together, our findings in conjunction with previous research suggest that studies aiming to detect social interaction in real time would benefit from including an accelerometer sensor among their passive sensing streams. 

Our results also suggest that EDA features, particularly the number of phasic peaks in skin conductance, are important for detecting when a person is actively engaged in social interaction. These findings are consistent with research that has found associations between heightened electrodermal activity and stress and anxiety (e.g., \cite{globisch_fear_1999, hickey2021smart}). However, the current findings differ from prior work suggesting that EDA features extracted from wrist-worn sensors may be unreliable in conversational settings \cite{milstein_validating_2020}. Indeed, it has been suggested that the Empatica E4's sampling rate of 4Hz may be too slow to accurately detect SCRs in the data \cite{braithwaite_guide_2013}. Although EDA metrics from wrist-worn devices may be limited in their ability to obtain objectively accurate readings, particularly when compared to laboratory-grade psychophysiological equipment \cite{kleckner_framework_2021}, they may still have some real-world utility when it comes to differentiating between contexts. In any case, replication of our findings with other devices is needed. 

To explore these findings in greater depth, we conducted clustering analyses on the two sets of tasks between which our models could reliably differentiate: alone vs. social and during vs. pre/post. The goal of these analyses was to examine whether participants exhibited shared or heterogeneous physiological responses when looking at the top-performing features in these tasks. Across both tasks, our results suggest that there is both considerable variability in physiological response patterns across individuals within the same context as well as important commonalities. For instance, there were five clusters representing different physiological response patterns across all social interactions, which suggests that there may be individual differences in ACC SD and SCR Peaksn during social interactions, but not to the extent that we would label these patterns as being entirely idiographic or person-specific. Accordingly, an important next step will be determining what, precisely, these clusters signify. For instance, it is plausible that the clusters represent state anxiety levels, and that socially anxious individuals exhibit different patterns of movement and skin conductance depending on how anxious they feel in the moment. Understanding these common physiological response patterns, and which feelings or behaviors they map onto, could aid in the deployment of JITAIs when and where they are most needed. 

\subsection{Limitations and Future Directions}
This study must be considered in light of its limitations and it is important to acknowledge that features of the study design may limit the generalizability of these findings to real-world contexts. First, the study was conducted virtually via Zoom and the within-person experimenter-manipulated conversation conditions (e.g., dyadic vs. group) all involved talking to strangers about benign topics, so the variety of interactions was limited. It is unclear to what extent these findings would hold under different conditions; when talking to friends vs. strangers or when interacting with people in person, for example, participants might rely less on gesturing as a communication tool and ACC SD could be less informative in differentiating between social and non-social contexts. Second, our sample was relatively homogeneous with respect to age (\textit{M} = 19.28, \textit{SD} = 1.91), sex (76.1\% female), race (67.4\% White), and ethnicity (89.1\% Non-Latinx/Hispanic). It is critical that we conduct future research with more diverse samples, particularly in light of findings that ostensibly "gold standard" psychophysiological measures like EDA were developed and tested with predominantly White participants and typical data cleaning procedures can systematically exclude non-White participants who exhibit lower reactivity \cite{bradford_whose_2022}. Third, there are inherent limitations with using wrist-worn devices to collect some psychophysiological data. For example, wristbands such as the Empatica E4 use dry electrodes (vs. wet electrodes prepared with a gel that enhances the electrical conductivity of the skin) and a low sampling rate that makes accurate measurement of skin conductance very challenging \cite{kleckner_framework_2021}. Moving forward, replication with different devices, larger and more diverse samples, and in real-world settings will be key. 

\subsection{Conclusion}
Despite the limitations, the present study advances our understanding of social context detection via passive sensing in several key ways. First and foremost, with the sensors used here we could reliably differentiate between social and non-social contexts but not between social interactions characterized by more subtle contextual differences. The accelerometer standard deviation and number of peaks in skin conductance response were the two most important individual features for differentiating between social and non-social contexts. Within these features, there was considerable variability in terms of physiological response patterns, but potentially meaningful clusters also emerged, providing some evidence for both individual differences and commonalities across participants. Taken together, these findings suggest that accelerometer and skin conductance sensors may be particularly important for identifying \textit{when} someone is in a social interaction, but not specific characteristics of that interaction. Importantly, even knowing whether or not someone is with others or not will be important for appropriately tailoring and deploying JITAIs. However, future work is needed to better detect characteristics of social interactions and to disentangle precisely what clusters of physiological response patterns in social interactions represent to further improve JITAIs. 

    
    \bibliography{main}



\end{document}